\documentstyle[12pt,epsfig,psfig,float]{article}
\date{\mbox{}}
\begin{document}
\title{\bf Nucleosynthesis in evolved stars with the
  NACRE\footnote{Nuclear Astrophysical Compilation of REaction rates
  (Angulo et al. 1999)} compilation }
%
%
%
%
\author{A.~Palacios$^1$, F.~Leroy$^2$,\\ C.~Charbonnel$^1$ and
  M.~Forestini$^2$}

\vspace{1cm}

\maketitle
\pagestyle{empty}
\noindent \normalsize $^1$ Laboratoire d'Astrophysique de Toulouse, CNRS UMR 5572,
France\\
\normalsize $^2$ Laboratoire d'Astrophysique de l'Observatoire de
Grenoble, CNRS UMR 5571, France\\
%
%
\def\bull{\vrule height .9ex width .8ex depth -.1ex}
\makeatletter
\def\ps@plain{\let\@mkboth\gobbletwo
\def\@oddhead{}\def\@oddfoot{\hfil\tiny\bull\quad
``Nucleosynthesis in evolved stars with the NACRE compilation'';
35$^{\mbox{\rm rd}}$ Li\`ege\ Int.\ Astroph.\ Coll., 1996\quad\bull}%
\def\@evenhead{}\let\@evenfoot\@oddfoot}
\makeatother
%
%
\def\beginrefer{\section*{References}%
\begin{quotation}\mbox{}\par}
\def\refer#1\par{{\setlength{\parindent}{-\leftmargin}\indent#1\par}}
\def\endrefer{\end{quotation}}
%
%

\vspace{0.5cm}

{\noindent\small{\bf Abstract:} 
Nucleosynthesis in evolved (RGB and AGB) low-mass stars is reviewed
under the light of the reaction rates recommended in the NACRE
compilation (Angulo et al. 1999). We use a parametric
model of stellar nucleosynthesis to investigate
the uncertainties that still exist nowadays on the nuclear data and to
give a critical point of view on
the resulting evolution of the chemical abundances. We discuss in
particular (i) the NeNa and MgAl modes of hydrogen burning in the
context of the chemical anomalies observed in RGB globular cluster
stars, (ii) the helium combustion in a thermal pulse of an AGB star.}
%
%
\section{Nucleosynthesis and abundance anomalies in RGB stars}
Proton capture nucleosynthesis inside the CNO, NeNa and MgAl loops, is
advocated to account for abundances anomalies observed at the surface
of GCRGs (Globular Cluster Red Giants), either in the context of the
primordial scenario or evolutionary scenario (see Sneden and
Charbonnel et al. in this volume). 
We focus here on the evolutionary scenario in order to
determine to what extent it can account for the observations from a nuclear
point of view. We present results for a $0.83 {\rm M}_{\odot}$, [Fe/H] =
-1.5 model which is typical of a RGB star in the globular cluster M13.\\
For our calculations we use the nuclear reaction rates recommended by the
NACRE consortium (Angulo et al. 1999). This compilation provides revised nuclear rates and, for the first time, gives the uncertainties
associated to these rates which we explored with a parametric
code of nucleosynthesis.\\  
\subsection{Influence of the NACRE reaction rates on the abundance profiles}
In fig.1, we present a comparison between the NACRE rates and earlier ones (Caughlan $\&$ Fowler 1988, Champagne et
al. 1988, Beer et al. 1991, Illiadis 1990, Gorres et al. 1989) for the
main nuclear reactions involving $^{24}{\rm Mg}$ and $^{27}{\rm
  Al}$. Shaded areas represent uncertainties associated to the NACRE
reaction rates. The uncertainty domain
can be quite large for some rates at the temperatures typical of the
Hydrogen Burning Shell inside an RGB star. This may
affect the abundance profiles of some elements in stellar interiors.\\

\begin{figure}[H]
  \centerline{
    \begin{minipage}{8cm}
      \epsfig{figure=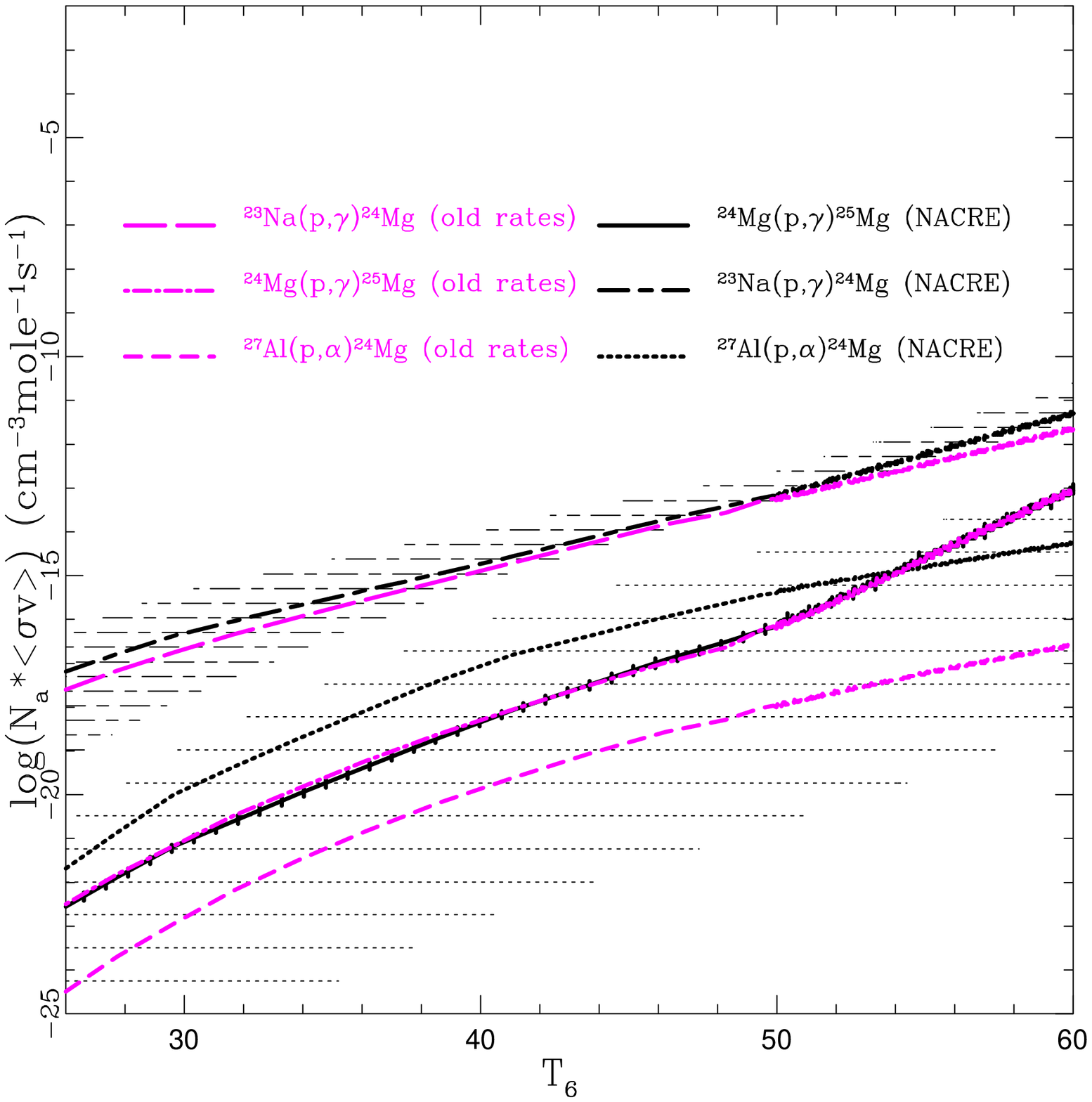,height=10cm,width=8cm}
    \end{minipage}%
    \begin{minipage}{8cm}
      \epsfig{figure=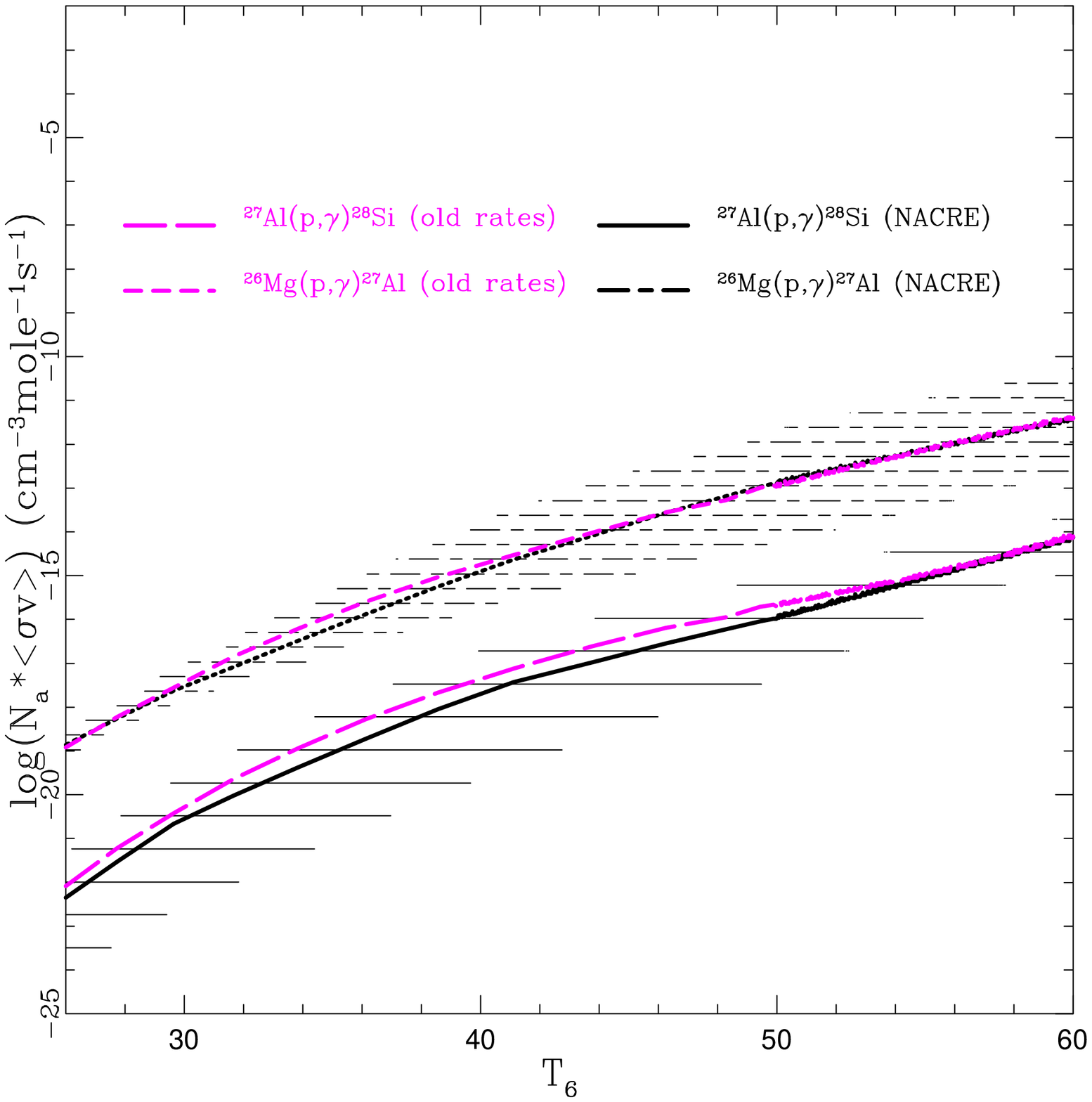,height=10cm,width=8cm}
    \end{minipage}}
    \caption{\footnotesize Comparison between the reaction rates recommended by NACRE
      and the ones from earlier sources (see references in
      text). Shaded areas represent uncertainty domains given by the
      NACRE compilation. The temperature domain corresponds to the
      regions surrounding the HBS between the RGB bump and tip in our model.} 
\end{figure}

Figure 2 presents the abundance profiles for CNO, NeNa and MgAl
elements between the bottom of the convective envelope ($\delta$M = 1)
and the base of the HBS ($\delta$M = 0), for a star typical of M13 at
the tip of the RGB. With the NACRE reaction rates (bold lines), the CNO abundance
profiles appear to be very similar to the ones obtained with the reaction rates
by Caughlan $\&$ Fowler (1988; thin lines). The
internal structure of the stellar models thus remains unaffected by using
either rates. This enables us to use a parametric code to study the
effects of the uncertainties on the reaction rates involved in the
 NeNa- and MgAl- loops, which appear to be very large in the relevant
 domains of temperature (fig.1).
The profiles of the carbon and oxygen isotopes appear to be
fairly well constrained. However quite large uncertainties remain in the
profiles (position and abundances) of $^{23}{\rm Na}$ and $^{25}{\rm
  Mg}$ which are mainly shifted toward the center or the external
layers, and of $^{21}{\rm Ne}$, $^{22}{\rm Ne}$, $^{26}{\rm Mg}$ and
$^{27}{\rm Al}$. For these isotopes, the maximum and minimum values of
their abundances at a given temperature can even change by one order
of magnitude.   

\begin{figure}[H]
  \centerline{
    \epsfig{figure=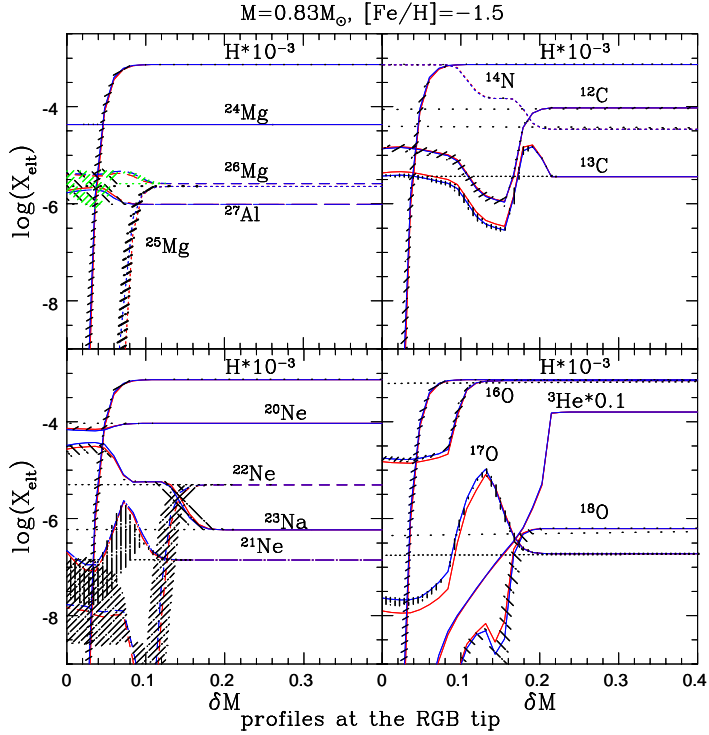,height=10cm}}
    \caption{\footnotesize Abundance profiles and uncertainties associated to the
    NACRE nuclear reaction rates. Bold lines correspond to the NACRE
    reaction rates and thin ones to earlier sources. $\delta$M is
    defined as : $\delta$M = ($M_r$ - $M_{core}$)/($M_{CE}$ - $M_{core}$)} 
\end{figure}
%
%
%
%

\subsection{Impact of the new rates on the evolutionary scenario}
Exploring these uncertainty domains, we now present the
maximum surface variations of sodium, oxygen, aluminium, etc..., that one can
expect to find within the framework of the evolutionary scenario.\\
In fig.3 and fig.4 we compare the nucleosynthesis predictions (bold
lines) with observations. From a nucleosynthesis point of view, enhancements of
sodium and depletion of oxygen observed at the surface of many GCRGs
can be reproduced assuming that some deep mixing process (which is not
taken into account in our model) connects the convective envelope and
the regions
where the abundance of helium has increased by less than 4$\%$. There,
within the more external sodium ``plateau'', hydrogen is not depleted
yet (see fig.2 and Charbonnel et al. in this volume).\\
\indent Figure 2 shows that no change in the $^{24}{\rm Mg}$ abundance is
noticeable with the NACRE rates in such a star as was already the case
with previous rates (Denissenkov et al. 1998, Lnager et al. 1993,
Langer $\&$ Hoffman 1993). $^{24}{\rm Mg}$ requiring higher temperatures to
capture protons (T $\sim$ 80.$10^6$ K), an 
enhancement of $^{27}{\rm Al}$ could only be due to $^{25}{\rm Mg}$
abundance decrease and to $^{26}{\rm Mg}$ increase, but not to
$^{24}{\rm Mg}$ destruction. 
On the other hand, without making any particular assumption on the
magnesium isotopic ratios in a M13 typical star (M = $0.83 {\rm
  M}_{\odot}$ and [Fe/H] = -1.5) (Shetrone 1996a and references therein), nucleosynthesis can not account for
the observed aluminium enhancements and the Na-O anticorrelation at
once as shown in fig.4. 
We confirm that the aluminium abundance anomalies observed by
Shetrone(1996) ($^{24}{\rm Mg}$ seems to be depleted in Al-rich stars) could be related to
primordial contamination. Testing various isotopic ratios for
magnesium, it turns out that with $^{24}{\rm Mg}/^{25}{\rm
  Mg}/^{26}{\rm Mg} \sim 29.4 / 58.8 / 11.8$, typical of intermediate
mass AGB yields (Forestini $\&$ Charbonnel 1997), one could
expect a more important but yet insuficient production of aluminium.\\
\indent In any case, new determinations of the Mg isotopic ratios in Al-rich
RGB stars are necessary to clarify the primordial effects.  

\begin{figure}[t]
  \centerline{
    \begin{minipage}{8cm}
      \epsfig{figure=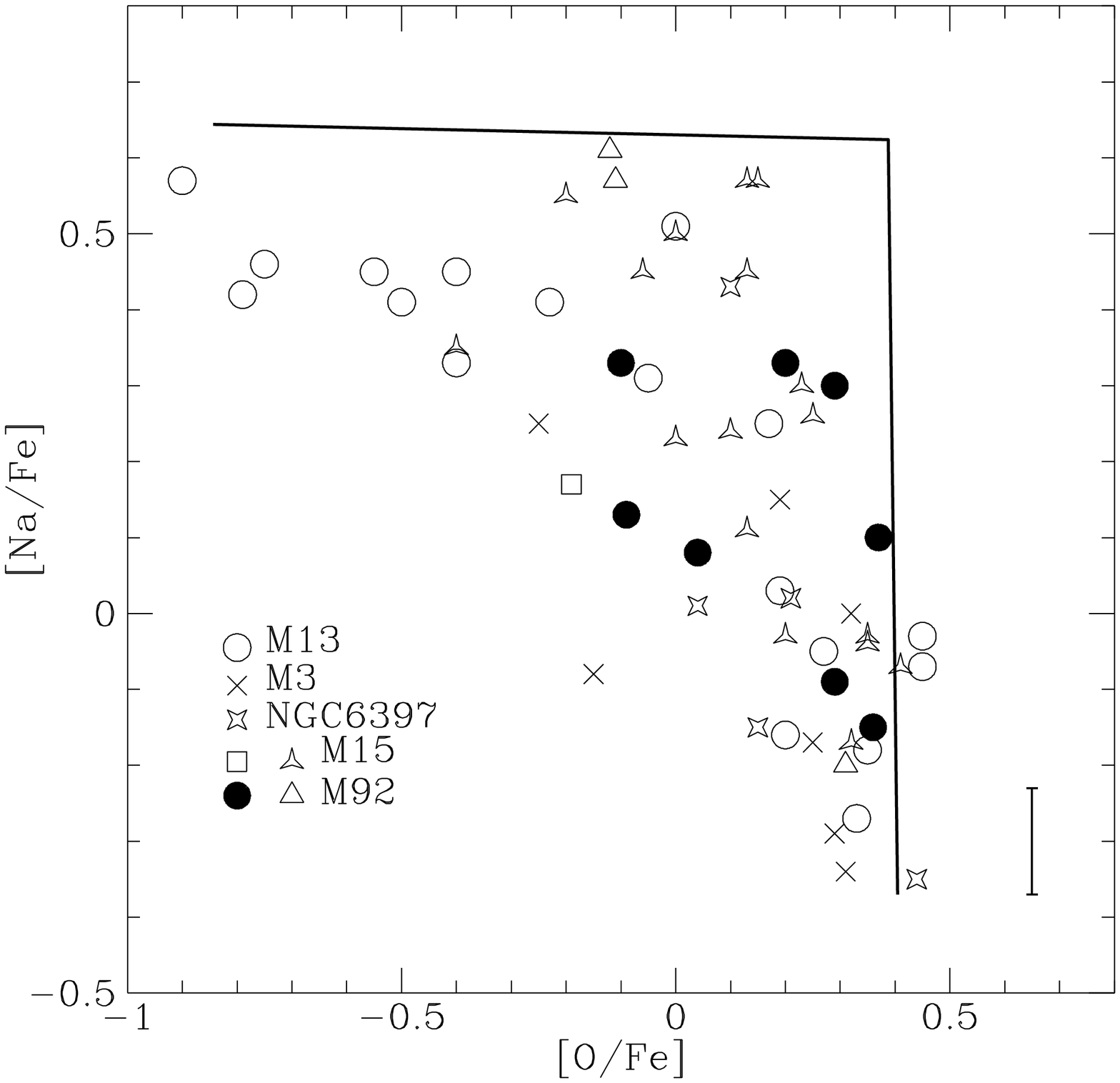,height=10cm,width=8cm}
      \begin{minipage}{7cm}
      \caption{\footnotesize O-Na anticorrelation in different globular
        clusters (see Charbonnel et al. in this volume for references). The bold line represents the variations
        obtained for a model typical of a RGB star in M13 (M = $0.83
        {\rm M}_\odot$, [Fe/H] = -1.5). The typical errorbar for
        Na abundance is also given.}%
     \end{minipage}%
    \end{minipage}%
    \begin{minipage}{8cm}
      \epsfig{figure=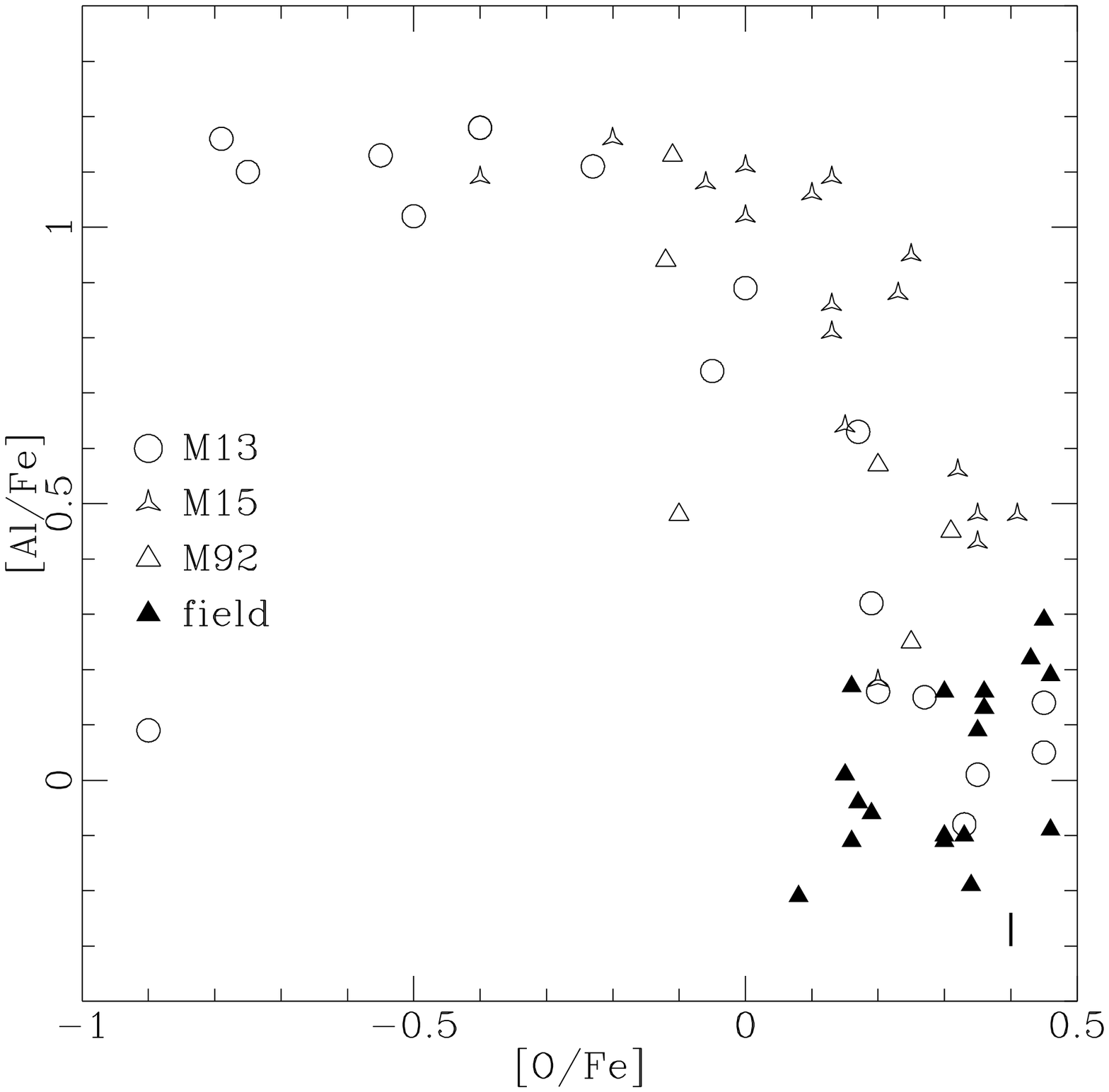,height=10cm,width=8cm}
     \begin{minipage}{7cm}
     \caption{\footnotesize O-Al anticorrelation in different
       globular clusters and in field stars. The bold line
       represents Al-variations within the more external sodium
       ``plateau'', which are compatible with the observed O-Na
       anticorrelation.\vspace{0.8cm}}
     \end{minipage} 
    \end{minipage} 
  }
\end{figure}

\section{Nucleosynthesis in an AGB star}
The NACRE consortium also provides reaction rates (and their lower and
upper limits) for the combustion of helium. We investigate their
influence on the evolution of the composition of the helium shell
surrounding the C-O core during the AGB phase. The NACRE rates allowed
us to show that the reactions responsible for the production of
$^{16}{\rm O}$ in intermediate mass AGB are well known and constrained. On the other hand, the rate of
the reaction $^{22}{\rm Ne}$($\alpha$,n)$^{25}{\rm Mg}$ is very
uncertain (uncertainty domain width : $\sim 10^3 - 10^4$ at ${\rm
  T}_8$ = 1 - 2); this prevents from giving good constraints to the
production of neutrons in massive AGB stars among other things.\\   
\indent We present here the results of the
$\alpha$-capture nucleosynthesis in the ``mean'' physical conditions
of a thermal pulse inside an AGB, obtained with a simplified parametric code (temperature and
density are constant). In particular, we describe in details the
variations of the abundances resulting of the two reaction chains
involving $^{14}{\rm N}$ and $^{18}{\rm O}$, and leading to the
production of $^{25}{\rm Mg}$ through $^{22}{\rm Ne}$.\\   
\indent During a thermal pulse, ashes from the HBS are melt
with helium at average temperature and density of ${\rm T}_8$ = 2.5
and $\rho = 10^4$ g/${\rm cm}^3$. Under these conditions
a very rich and unique nucleosynthesis occurs, as shown in fig.5.

We focus on the three following reaction chains:

\begin{equation}
^{16}{\rm O}(\alpha,\gamma)^{20}{\rm Ne}(\alpha,\gamma)^{24}{\rm Mg}
\end{equation}
\begin{equation}
^{14}{\rm N}(\alpha,\gamma)^{18}{\rm F}(\beta)^{18}{\rm
  O}(\alpha,\gamma)^{22}{\rm Ne}(\alpha,n)^{25}{\rm Mg}
\end{equation}
 and 
\begin{equation}
^{14}{\rm N}(\alpha,\gamma)^{18}{\rm F}(\beta)^{18}{\rm
  O}(p,\alpha)^{15}{\rm N}(\alpha,p)^{19}{\rm F}(\alpha,n)^{25}{\rm
  Mg}
\end{equation}
  We can notice that all the reactions involve the capture of an
  $\alpha$-particle, except $^{18}{\rm O}$(p,$\alpha$)$^{15}{\rm
  N}$. In this case, we will show that (n,p)
  reactions provide the flux of protons necessary to make the
  (p,$\alpha$) channel efficient for the destruction of $^{18}{\rm O}$
  in the third chain.

\begin{figure}[t]
\caption{\footnotesize Main nuclear reaction fluxes occuring
      during a thermal pulse by the light of the NACRE data base. The
      bolder the arrows, the more important the flux.}
\end{figure}

\vspace{0.5cm}

{\large\bf{Proton and neutron sources during a thermal pulse}}
\vspace{0.5cm}

In fig.5, we present the main nuclear reaction fluxes during the
combustion of helium inside the thermal pulse. The 3$\alpha$ reaction produces a large amount
of $^{12}{\rm C}$, which, together with $^{13}{\rm C}$ (which comes from
H-burning ashes), leads mainly to the synthesis of $^{16}{\rm
  O}$. Indeed, the reactions responsible for the production of $^{20}{\rm Ne}$ and
$^{24}{\rm Mg}$ have sufficiently low rates to keep the abundances of
these elements unchanged (reaction chain (1)).
The reaction $^{13}{\rm C}$($\alpha$,n)$^{16}{\rm O}$ is the
main source of neutrons in the medium at the beginning of the
helium flash, as shown in figure 6. This production of
neutrons is directly correlated to the abundance of protons through
the reaction $^{14}{\rm N}$(n,p)$^{14}{\rm C}$ at this point in the
He-burning phase (see fig.6).\\
Reaction chain (2) is generated by a ``$^{14}{\rm N}$ reservoir'',
which has been built up during the hydrogen combustion through the CNO
cycle. Each element involved in this chain is produced and then
destroyed by the capture of an $\alpha$-particle.
Concerning $^{18}{\rm O}$, however, we can notice that the
(p,$\alpha$) channel, even if narrower than the ($\alpha$,$\gamma$)
channel, can not be neglected (reaction chain (3)). This is due to the
presence of protons in the medium, that we can impute to both
$^{14}{\rm N}$(n,p)$^{14}{\rm C}$ and $^{26}{\rm Al}^{\rm
  g}$(n,p)$^{26}{\rm Mg}$ reactions. Indeed, an analysis of the fluxes
derived from the NACRE reaction rates shows that these two reactions
provide an equivalent amount of protons in so far the abundance of
$^{14}{\rm N}$ is not negligible. We have already mentionned
$^{13}{\rm C}$($\alpha$,n)$^{16}{\rm O}$ as main supplier of neutrons
in the early stages of He-burning, and we can add  $^{22}{\rm
  Ne}$($\alpha$,n)$^{25}{\rm Mg}$, which takes over at more advanced
stages (see fig.6).
\enlargethispage{+2\baselineskip} 
\begin{figure}[t]
  \centerline{
      \epsfig{figure=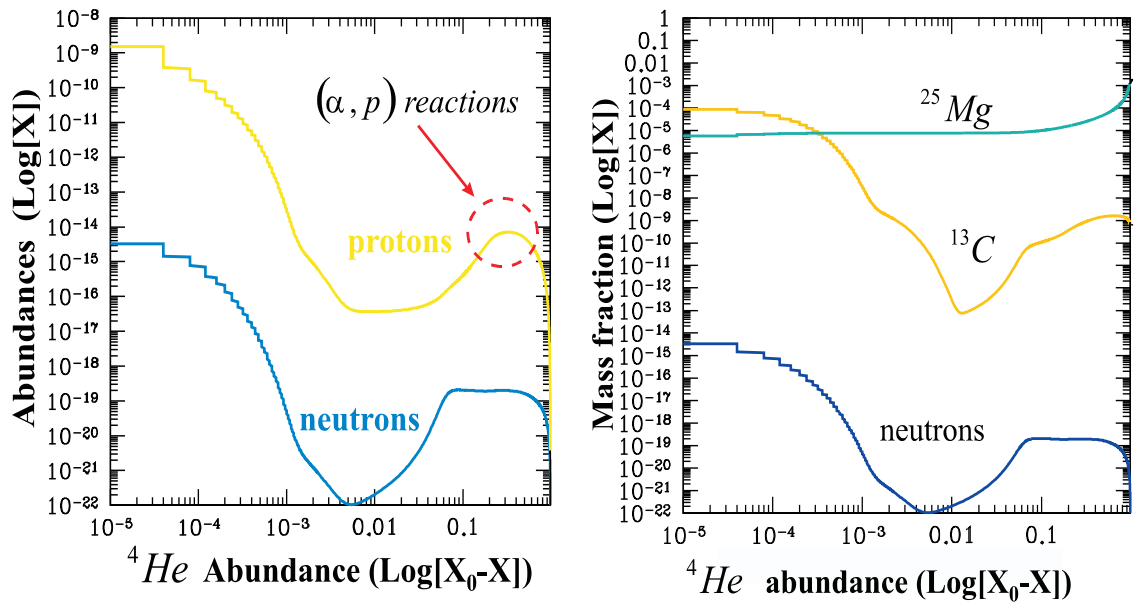,height=8cm,width=15cm}}
      \caption{Correlation between the abundances of protons and
      neutrons (left pannel) and neutrons production (right pannel) as
      functions of the amount of $^4{\rm He}$ that has been burnt.}
\end{figure}

\beginrefer
\refer Angulo C., Arnould M., Rayet M., et al. (the NACRE
collaboration), 1999, Nucl. Phys., A656, 3

\refer Beer H., Voos F., Winters R.R., 1991 preprint

\refer Caughlan G. and Fowler W.A., 1988, At. Data Nucl. Data Tables,
40, 207

\refer Champagne A.E., Cella C.H., Konzes R.T., Loury R.M., Magnus
P.V, Smith M.S., Mao Z.Q., 1988, Nucl. Phys. A487 

\refer Denissenkov P.A. and Denissenkova S.N., 1990, Soviet
Astron. Lett., 16, 275

\refer Denissenkov P.A., Da Costa G.S., Norris J.E. and Weiss A., 1998,
A$\&$A, 333, 926-941

\refer Forestini M. and Charbonnel C., 1997, A$\&$A Suppl. Ser., 123,
241

\refer Gorres J., Wiescher M., Rolfs C., 1989, ApJ 343, 365

\refer Illiadis C., 1990, Nucl. Phys. A512, 509

\refer Langer G.E., Hoffman R. and Sneden C., 1993, PASP, 105, 301-307  

\refer Langer G.E. and Hoffman R., 1995, PASP, 107, 1177-1182

\refer Shetrone M.D., 1996a, AJ 112, 1517

\refer Timmermans R., Becker H.W., Rolfs C., Schroder U., Trautvetter
H.P., 1988 Nucl. Phys. A447, 105

\endrefer           
\end{document}